# 2D Functionalized Ultra-thin semi-insulating $CaF_2$ layer on the Si(100) surface at low temperature (9K) for molecular electronic decoupling

Eric Duverger[1], Anne-Gaëlle Boyer[2], Hélène Sauriat-Dorizon[2], Philippe Sonnet[3,4], Régis Stephan[3,4], Marie-Christine Hanf[3,4], Damien Riedel[5*]

[1]*Institut FEMTO-ST, Univ. Bourgogne Franche-Comté, CNRS, 15B avenue des Montboucons, F-25030 Besançon, France.*

[2]*Institut de Chimie Moléculaire et des Matériaux d'Orsay (ICMMO), CNRS, Univ. Paris Sud, Université Paris-Saclay, F-91405 Orsay, France.*

[3]*Institut de Science des Matériaux de Mulhouse, Université de Haute Alsace, CNRS, IS2M UMR 7361, 68057 Mulhouse, France,*

[4]*Université de Strasbourg, F-67000 Strasbourg, France.*

[5]*Institut des Sciences Moléculaires d'Orsay (ISMO), CNRS, Univ. Paris Sud, Université Paris-Saclay, F-91405 Orsay, France.*

*Corresponding author: damien.riedel@u-psud.fr*

**Abstract:**

**Our ability to precisely control the electronic coupling/decoupling of adsorbates from surfaces is an essential goal. It is not only important for fundamental studies in surface science, but also in several applied domains including for example miniaturized molecular electronic or for the development of various devices such as nanoscale bio-sensors or photovoltaic cells. Here, we provide atomic scale experimental and theoretical investigations of a semi-insulating layer grown on a silicon surface via its epitaxy with $CaF_2$. We show that, following the formation of a wetting layer, the ensuing organized unit-cells are coupled to additional physisorbed $CaF_2$ molecules, periodically located in their surroundings. This configuration shapes the formation of ribbons of stripes that functionalize the semiconductor surface. The obtained assembly, having a monolayer thickness, reveals a surface gap energy of ~ 3.2 eV. The adsorption of iron-tetraphenylporphyrin molecules on the ribbons of stripes is used to estimate the electronic insulating properties of this structure via differential conductance measurements. Density functional theory (DFT) including several levels of complexity (annealing, DFT+U and non-local van der Waals functionals) are employed to reproduce our experimental observations. Our findings offer a unique and robust template that brings an alternative solution to electronic semi-insulating layer on metal surfaces such as NaCl. Hence, $CaF_2$/Si(100) ribbon of stripes structures, which lengths can reach more than 100 nm, can be used as a versatile surface platform for various atomic scale study of molecular devices.**

1. INTRODUCTION

The growth control, as well as the properties of electronic insulating layers, on various materials surfaces is of key importance for many domains, not only including electronic devices[1]. This topic has thus been the object of a large amount of fundamental[2] and applied investigations[3]. It is still impacting various research fields targeting both applied and fundamental physics in which flexible opto-electronic devices[4,5,6], sensors[7,8] or biological functions are still feeding new paradigms[9,10].

In many research fields, the growth of mesoscopic devices having relatively thick dielectric layers made for example of $SiO_2$ on semiconductors or $Al_2O_3$ on metals surfaces (MS) often reach wideness of few nanometers. This renders their fabrication seldom repeatable at the atomic scale and may well also present conducting leaks[11].

At lower scale, the approach is different since current leakage can be exploited for specific tunnel transport in molecular scale devices[12,13]. In the context of molecular electronics, ultra-thin insulating layers made of a few atomic crystallographic scale coats have already been largely investigated. Several well-known materials such as NaCl[14] as well as $Al_2O_3$[15,16] have demonstrated efficient electronic decoupling properties of single atoms or molecules on Cu, Au or Ag surfaces. However, on metal surfaces, few layers of dielectric material are necessary to reach electronic insulating properties with surface energy gaps reaching ~ 3 eV. Hence, at this scale, the scanning probes techniques allow investigating ways to tune their electrical insulating properties, in particular for the study of photonics or electronic devices[17].

Yet, on semiconductor surfaces, except for $SiO_2$ dielectric, the growth control of ultrathin semi-insulating layers remains barely studied and therefore faintly understood, especially at atomic scale thickness. This arises mainly because of the anisotropy of semiconductor surfaces reconstructions that may prevent a good 2D lattice matching with the electronic insulating material. Additional surface passivation techniques, e.g. with elements like hydrogen[18] or via the use of molecular layers[19] have also been explored but the obtained surface band gap energy is often reduced to 1-2 eV.

As the development of active devices at smaller scales remains a quest in many domains including molecular electronic[20] or nanoscale magnetism[21] for which a thorough control of the atomic scale environment is decisive, it is particularly vital to determine the atomic and electronic structures of a semi-insulating layer able to support a molecular or atomic assembly.

To the best of our knowledge, ultrathin insulating layers using $CaF_2$ is the sole alternative to reach efficient electronically insulating properties of semiconductor surfaces. It offers a clear alternative to NaCl layers on metal surfaces. Alkaline earth fluorides such as $BaF_2$ or $SrF_2$ have also been explored as candidates on silicon surfaces but with reduced insulating efficiencies[22]. On Si(111) surfaces, the structure of the ultrathin $CaF_2$ reconstruction has been often investigated and exhibits a relatively weak surface gap of ~1.7 eV[23,24,25] and is still an attractive candidate for mesoscopic MISFET (Metal Insulator Semiconductor Field Transistors) structures[26]. At smaller scale, the epitaxial growth of $CaF_2$ on Si(100) surfaces exhibits ultrathin insulating anisotropic reconstructions forming ribbons of stripes that offer additional functionalities for organizing adsorbates[27]. While the electronic structure of the $CaF_2$/Si(100) has been investigated with mesoscopic techniques[28], it is of crucial importance for the understanding of their electronic structure to investigate their atomic scale formation[29]. Yet, a detailed and complete description of the precise atomic and electronic structure of ultrathin $CaF_2$ insulating ribbons of stripes on Si(100) is lacking, as well as the characterization of the interaction of this type of semi-insulating layer with molecular adsorbates.

In this article, we investigate experimentally and theoretically the atomic scale structure of $CaF_2$ ribbons of stripes resulting of the growth of ~ 1.3 monolayer of $CaF_2$ on the bare Si(100) surface. The scanning tunneling microscope (STM) topographies and spectroscopic measurements obtained at 9 K provide a precise picture of the stripes structures. As well, it provides the relative positioning of the ribbons with respect to the silicon dimer rows and the ensuing wetting layer (WL) that is initially on the bare silicon. Our investigations also provide a complete characteristic of the electronic insulating properties via an estimation of the $CaF_2$ layer surface gap energy. The surface gap is considered here as different from the bulk gap energy and defined as the energy difference between the upper part of the valence band and the lower part of the conduction band for the first atomic layers of the surface, as previously studied[35]. In order to determine precisely the atomic structure of the stripes and confirm our experimental findings, we have used the density functional theory (DFT) with annealing procedure and calculated STM images. Additionally, the DFT-Hubbard method is used to provide additional insights on its electronic structure and allows to estimate the surface energy gap with the projected density of state (PDOS) information. Finally, in order to estimate the quality of the electronic decoupling, we use STM analysis combined with van der Waals functionals within DFT method to describe the interactions of a single iron-tetraphenylporphyrin (FeTPP) molecule with a ribbon of $CaF_2$ stripe. The ensemble of our results indicates that this two- dimensional electronic insulating material allows a strong electronic decoupling of organic or inorganic adsorbates.

## 2. EXPERIMENTAL SECTION

### 2.1. Synthesis and epitaxial preparation of the $CaF_2$/Si(100) surfaces.

The characterization of the ultrathin insulating layer is performed with a low temperature (9 K) PanScan STM head from CREATEC. The Si(100) samples (n-doped (As), $\rho = 5\Omega.cm$) are cleaned in ultra-high vacuum using repetitive annealing cycles consisting of a rapid temperature increasing up to 1100 °C followed by a slow temperature ramp down to 600 °C. The reconstructed Si(100) surface is then exposed to a flux of $CaF_2$ molecules evaporated from an effusion cell, previously calibrated with a quartz balance, heated at 1065 °C, while the Si sample is kept at a temperature of 720 °C. The $CaF_2$ exposure, lasts for 120 s for a total coverage of 1.3 monolayers. During the $CaF_2$ exposure the pressure in the preparation chamber is kept below 2 x $10^{-10}$ torr. Following this procedure, the sample is cooled down to 12K and introduced in the STM chamber for analysis. Note that the phase diagram of the $CaF_2$ epitaxy on the Si(100)-2x1 is noticeably different than the one for the Si(111)-7x7 reconstruction, leading to significant variation of substrate growth temperatures[25,28].

### 2.2. Molecule adsorption and characterization

Following the experimental characterization of the bare $CaF_2$ ultrathin insulating layer, the sample is extracted from the STM and kept at low temperature (12 K) to be exposed to a flux of iron tetra-phenylporphyrin (FeTPP) molecules via the heating of a PBN Knudsen cell at ~ 230°C. The fixed exposure time (2 s) leads to a coverage of ~ 0.4 monolayer. Note that in the following, we will consider that the chlorine apical ligand initially fixed on the Fe atom of the FeTPP molecule is removed due to the evaporation process.

### 2.3. Theoretical methods and means.

We have used the Spanish Initiative for Electronic Simulations with Thousands of Atoms (SIESTA)[30,31] package and combined simulated annealing optimization technique to find the ground state structure of the $CaF_2$ stripes obtained on the Si(100) surface. The Perdew–Burke–Ernzerhof (PBE) Generalized Gradient Approximation (GGA) for the exchange-correlation density functional was employed to determine the total energies[32,33]. During the annealing simulations, the atoms motions follow the Verlet algorithm velocities with a time step of 1 fs[34]. The system is then raised at a high temperature (1000 K) and slowly cooled down to 4 K in about 20000 steps. The relaxation of the entire system ($CaF_2$ molecules + WL) is performed with a grid of 9 x 9 x 3 k-points giving a total of 135 k-points in the irreducible Brillouin Zone. Then, the following standard set of parameters are chosen such that the self-consistency mixing rate is fixed to 0.1, the maximum force tolerance is set to 0.02 eV.Å$^{-1}$ and the mesh cut-off is adjusted at 400 Ry. The calculations are performed without spin polarization and a basis set of localized atomic orbitals (double-ζ plus polarization functions) with norm-conserving pseudopotentials are employed. The self-consistent cycles were stopped when the variations of the total energy per unit-cell and the band structure energy are both lower than $10^{-4}$ eV. The final optimized unit-cell (i.e., 23.20 x 7.78 x 44.50 Å$^3$) contains 156 atoms (i.e. 4 Ca, 120 Si, 24 H and 8 F) and 10 Si layers in order to take into account subsurface interactions. For the calculations performed on the sole FeTPP molecule in the gas phase, we use the Gaussian V09 simulation package. The structure of the molecule is relaxed and the Kohn-Sham energies are computed with the HSE hybrid functional (HSEh1PBE). The C and H atoms are described with the 6-31G* basis set while the 6-311G* basis set is chosen for the Fe and H atoms.

### 3. RESULTS AND DISCUSSION

The epitaxy of the Si(100) surface by $CaF_2$ molecules follows a Volmer-Weber (VW) growth law. The initial stage of the epitaxy in which the formation of a WL occurs, previously studied by our team, was investigated at a low $CaF_2$ coverage (~ 0.3 monolayer)[29]. The WL formation implies that the silicon surface is periodically etched during the adsorption/dissociation of the $CaF_2$ molecules to form a (2x3) unit-cell as recalled in Fig. 1a. The WL forms patches appearing as dark corrugations in Fig. 1b. Inside these patches, each unit-cell is oriented along the silicon dimer rows [110] direction and their relative direction can be identical or head-to-tail[29]. Increasing the amount of $CaF_2$ molecules during surface exposition results in observing the second step of the VW process leading to the formation of ultra-thin ribbons of stripes, as it can be observed in Fig. 1b. At this stage, it is unknown if the additional $CaF_2$ molecules are partially dissociated or if they are kept unspoiled to form the stripes. Additionally, the electronic insulating character of these stripes in not yet demonstrated[27] and the precise atomic scale structure of the ribbons of stripes remains unknown.

A careful look in Fig. 1b shows that a ribbon is composed of several parallel stripes oriented perpendicularly to the unit-cells of the WL, i.e. the silicon dimer rows. Note that the stripes are surrounded by either residual bare silicon dimer rows or spots of WL[29]. The averaged distance between each stripe is 11.7 ± 0.3 Å. Paying attention on each stripe, one can identify that they appear to be composed of a series of twisted chevrons. As we emphasized, the ribbon of stripes seems to be sitting on the WL. One of the important parameter for the understanding of the growth of a $CaF_2$ semi-insulating layer is thus its relative orientation between the WL unit-cells.

To investigate the inner structure of a stripe, we have enhanced the measured height variations of a portion of a ribbon via a three dimensional representation of the STM topography. It reveals a measured corrugation between two stripes of ~ 0.5 Å (Fig. 2a, see also the supplementary Fig. S1). The stripes appear to be composed of left and right chevrons, each of them made of shifted bright features. The red dotted lines in Fig. 2a locate the chevrons maxima. The zoom on a single stripe shown in Fig. 2b estimates precisely the separation between the left and right chevrons in the order of ~ 1.4 Å.

The atomic scale structure of a ribbon of stripes can also be compared with the relative positions of the silicon atoms surrounding the stripes as shown in Fig. 2c. By locating the direction of the silicon dimer rows (horizontal red dotted lines) and the maximum density of state (DOS) intensity at the stripes (horizontal blue dotted line), one can determine that both have the same periodicity (7.8 ± 0.2 Å and 7.9 ± 0.3 Å, respectively). One can also notice that the lateral distance between these two directions is similar to the size of a silicon dimer row width (i.e. 2.6 ± 0.2 Å). The STM image shown in Fig. 2c allows to locate the position of each silicon dimers along the bare Si(100) surface, via the identification of the silicon back-bonds (BB) directions [35] by vertical dotted white lines (Fig. 2c). A detailed explanation of the positioning of the back-bonds is recalled in Fig. S2 and are observed as darker spots of DOS separated by brighter lines in the back-bond rows. Combining these two perpendicular directions (red horizontal and white vertical dotted lines) forms a grid that can be extended and superimposed over the ribbon of stripes area. From this, we can locate the relative positions of the oriented WL unit-cells in the area of the ribbon of stripes, as if they were part of this structure. The direction of the WL unit-cell is here chosen via a careful comparison of the surrounding of the ribbon of stripes[29]. This assumption reveals how the substitutional Ca atoms of the WL can be aligned along the stripes (light green balls of the cells). Coherently to the second Volmer-Weber growth phase, when additional $CaF_2$ molecules are evaporated over these areas, they are not likely to dissociate during their adsorption, as for the formation of the WL, but rather interact with the existing structure as an interstitial element. Hence, to form the observed ribbon of stripes oriented in the [1-10] direction, it is consistent to consider that the additional $CaF_2$ molecules are mostly located in between each WL unit-cell and most probably within the substitutional Ca atoms of each cells.

To understand how the additional $CaF_2$ molecules react with the organized WL, we have performed DFT calculations on a silicon slab having a surface area made of 4 WL unit-cells (see section 2.1 for details). We have thus based our investigation by adding additional $CaF_2$ molecules near two consecutive Ca restatoms of two WL unit-cells. In a first step, the entire relaxed geometry of the slab is obtained from DFT calculation by optimizing the energy of the initial large structure of the wetting layer for both Si atoms and surface unit-cell. In a second step, the requested number of $CaF_2$ molecules (four) is added on the designed slab and we then proceed to the annealing and relaxation of the whole system (see Fig. S3 for further details). To simulate the experimental annealing and reproduce the dynamics of the ribbon formation, the random initial velocities, corresponding to a temperature of 1000 K, are assigned to all the atoms of the surface by using Maxwell–Boltzmann distribution.

The final relaxed structure is shown in Fig. 3a-c. Figs. 3b and 3c show the two side views of the slab as described in the top view in Fig. 3a. It is interesting to notice that, following the relaxation, the substitutional Ca atoms are not anymore centered at the missing Si dimer of the WL unit-cell but shifted towards the two F atoms (F1) that passivate the neighbor Si dimer. Our calculations show that the stripes shape observed in Fig. 2 are formed via the intercalation of additional $CaF_2$ molecules (Ca1 and F2 in Fig. 3a) along with the Ca of the WL unit-cell. Here, the interactions of the Ca atoms with the four F atoms (i.e. two chemisorbed on the Si dimer of the WL cell and the two from the $CaF_2$ molecule) are significant for the stabilization of the entire stripe structure. As a consequence, the distance that separates two consecutive Ca atoms along a stripe is ~1.4 Å while the side views in Figs. 3b, 3c reveal a height variation between two successive Ca atoms of ~ 0.7 Å. Hence, the calculated periodicity between two repeated stripes is 11.7 Å (Fig. 3c). Logically, the length separating two chevrons (i.e., two substitutional Ca atoms) is ~7.6 Å. For the sake of clarity, we have also compared the calculated local density of states (LDOS) distribution of the simulated surface with the experimental STM images (Figs. 3d and S4). All these data are in very good agreement with the experimental measurements reported in Figs. 1 and 2.

It is now essential to examine the electronic structure of the ribbon of stripes. For this, we have measured the averaged differential conductivity by acquiring various dI/dV curves at different positions across a ribbon of stripes. The result of the averaged measurements is presented in Fig. 4a. The surface gap energy can be deduced by defining the asymptotic intersections at the valence and conduction bands edges with the zero conductance plateau. The crossing values worth -1.13 ± 0.10 V and 2.09 ± 0.10 V, giving a total surface gap energy of 3.22 ± 0.20 V. This experimental value can be compared with a theoretically estimate of the surface gap energy of the semi-insulating layer performed via a specific calculation. We used the DFT+U approach method implemented in SIESTA based on the Dudarev et al. scheme[36] that combines the two parameters U and J to produce an effective Coulomb repulsion energy defined by $U_{eff} = U - J$. In order to retrieve the experimental surface gap energy, given the intrinsic difficulty in determining the ab-initio value of $U_{eff}$, we have computed the surface gap evolution as a function of $U_{eff}$ varying from 2 eV to 30 eV. Although the implemented DFT+U approach in SIESTA allows defining independently the U and J parameters for each atomic shell, we have chosen to use the same $U_{eff}$ value for the Ca, Si and F atoms. Each PDOS curve is defined by 500 points spanning over the [-20 eV to 10 eV] energy window from either side of the Fermi energy value. The smearing energy used for the PDOS curves is fixed at 0.3 eV. The ensuing evolution of the calculated surface gap as a function of the tested $U_{eff.}$ values is reported in Fig. 4b. The surface gap remains negligible (ie < 200 meV) for $U_{eff.}$ varying between 0 eV and 4.45 eV and then increases almost linearly until $U_{eff.}$ reaches 26.45 eV to finally decreases sharply. Therefore, we deduced that the most accurate value of $U_{eff.}$ reproducing the measured surface gap energy on the ribbon of stripes is 26.45 eV. An example of a PDOS curve at this $U_{eff.}$ value is given in Fig. 4c with the detailed contribution of the F, Si and Ca atoms. Note that there are no dopants atoms in the considered slab for the calculations. This can explain why the experimental and theoretical Fermi energy are not the same in Figs. 4a and 4c. It is relevant to observe that the Hubbard correction in the DFT+U method induces energy shifts of the occupied states towards lower energies, rendering the estimation of the surface gap energy relatively accurate. From the subsequent total PDOS of the considered example reported in Fig. 4c, it is possible to estimate a calculated surface gap energy of 3.1 eV (Fig. 4d).

The charge states of the various elements forming the stripes are analyzed via the Bader scheme[37,38]. The initial effective charge for the valence electrons of the Ca atoms is 2e. After relaxation (without Hubbard correction), the charges of the residual Ca of the additional $CaF_2$ molecules (Ca1 in Fig.3a) are at 0.891e and 0.897e, respectively, thus having lost 1.109e and 1.103e. When the Hubbard correction is applied for fitting the experimental surface gap energy, the Ca effective charge is worth 0.173e and 0.171e, leading to an even stronger charge loss (i.e. 1.827e and 1.829e, respectively). Concerning the substitutional Ca atoms of the WL (Ca(2) in Fig.3a), their charge states are slightly different (0.768e without Hubbard correction and 0.202e with correction) but show the same trends (+1.80e). The charge variation of the F atoms changes in the opposite way, as one would expect, mimicking the $CaF_2$ ionic crystal. Here, the F atoms have an initial charge of 7 electrons. With Hubbard correction, the F atoms attached on the Si dimers nearby the substitutional Ca2 atoms (F1 in Fig. 3a) win negative charges to reach 7.59e whereas the charge of the F atoms forming the $CaF_2$ molecule is slightly higher with 7.82e. This reveals that the F atoms of the $CaF_2$ molecules added on the slab remain strongly bonded to the Ca atom and form an ionic bonding network with the substitutional Ca atom of the WL layer. Note that our calculated effective charge of a $CaF_2$ molecule are similar to former work, i.e. a loss of ~1.80e for $Ca^+$ and a gain of ~ 0.88e for $F^-$, respectively[39]. Therefore, we can conclude that the simulated stripes structure that takes into account the Hubbard correction fits correctly our experimental observations and shows a good trend of the electronic structure of the insulating layer. A detailed description that summarizes these data is given in the supplementary Fig. S5. It is important to underline that the relatively high energy of $U_{eff.}$ needed to reproduce the experimental surface gap energy of ~ 3.2 eV is not so surprising considering the large amount of computational parameters that may affect this tuning value, and the lack of initial localization in the exchange correlation functional that traduce Coulomb interactions. One can however observe that the Hubbard correction affect the charge state of the Ca and F atoms of the insulating layer in a way where the ionic properties of the Ca-F ionic bonding is enhanced. However, as shown in Fig. S6, the geometry of both type of Ca atoms (Ca1 and Ca2) of the insulating layer is only weakly affected (less than 2 %) by the Hubbard correction $U_{eff.}$ = 26.45 eV. Note that other approaches using ab-inito method such as unrestricted Hartree Fock may also give interesting results but are beyond the scope of the present paper[40].

To test the electronic semi-insulating properties of the ribbon of stripes, we have studied how a FeTPP molecule is perturbed when adsorbed on the epitaxial surface. As previously investigated, this molecule shows, on this surface, two main conformations having a mirror symmetry[20]. Only one of these configurations is considered here (Figs. 5a, 5b). While the molecule may be initially considered as physisorbed, its electronic structure on the semi-insulating surface is clearly unknown.

The ensuing conformation of a FeTPP molecule adsorbed on a stripe is detailed in Fig. 5b. Further details of the considered slab for this calculation are provided in Fig. S8. In order to take into account Van der Waals interaction between the substrate and the molecule, relaxation of the overall system was performed with double-ζ plus polarization functions orbitals with VdW-DF functional (DRSLL)[41,42]. From these calculations, the relative height of the FeTPP molecule compared to the surface can be deduced. For example, the Fe atom of the FeTPP (center of the molecule) is located at 5.4 Å from the surface. The nearest atoms of the molecule from the surface are hydrogen atoms of the phenyl rings located along the stripe with a typical distance of 2.9 Å. These values confirm the physisorbed character of the adsorbate. The calculated STM image of a FeTPP molecule adsorbed in this conformation is shown in Fig. 5d and presents good agreements with the experimental one (Figs. 5a, see also the Fig. S7 for additional details). To highlight the perturbations of the surface on the molecule or vice versa, the Bader charges are computed on the whole slab. Here, the calculated total charge transfer between the substrate and the molecule is relatively low and worth 0.053e. In order to further highlight the interaction between the molecule and the substrate, we have computed the variations of the LDOS in the slab. It is obtained when the LDOS is computed for the substrate ($LDOS_{sub.}$) and the molecule ($LDOS_{Mol.}$) separately, in one first step. The difference $\Delta(LDOS)$ is then achieved, as a second step, via the subtraction of the total $LDOS_{Tot.}$ computed when the molecule is adsorbed on the surface with the two other values ($\Delta(LDOS) = LDOS_{Tot.} - LDOS_{sub.} - LDOS_{Mol.}$). The results of the charge densities differences are shown in Figs 5e and 5f. The red and blue iso-surfaces represent positive or negative differences of the $\Delta(LDOS)$ values, respectively. One can firstly observe that the $\Delta(LDOS)$ is globally weak since the selected isodensity values is in the range of $10^{-4}$ e/Å$^{-3}$. Furthermore, the LDOS variations are mainly concentrated on the phenyls of the FeTPP molecule and especially on the two phenyls located along the insulating stripe (see black arrow in Fig. 5e for Ph1 and Ph2). The concentration of $\Delta(LDOS)$ on the upper and lower phenyls is further detailed in Fig. 5f, which mainly reveals that the lower part of the molecule is perturbed by the surface. A detailed analysis of the molecule-surface interactions reveals that after the FeTPP molecule adsorption, one of the lower carbon atoms from Ph2 can lose up to 0.09e, while the carbons atoms of Ph1 are less impacted (maximum gain of 0.008e). The charge state of the hydrogen atoms from the two considered phenyls are not impacted at all for Ph1 and only slightly changed for Ph2 (gain of 0.02e). As a consequence, the underlying Ca atoms below Ph2 and Ph1 lose both ~ 0.05e. The F atoms in the vicinity of these phenyl rings are also gently impacted with a maximum charge gain of 0.03e. Considering the relatively long distance separating the lowest hydrogen atom of the FeTPP molecule to the surface, these small charge perturbations are coherent with van der Waals forces interactions. A detail analysis of the charge state distributions of the FeTPP molecule in interaction with the surface is provided in Fig. S9. It is important to underline that the charge variations presented in the case of the bare insulating surface (Figs. 3 and S4) are slightly different due to the fact that the Hubbard correction is not taken into account in the case of the calculations of the molecule on the surface.

Similarly to what has been done with NaCl insulating layers in the presence of charged adsorbates[43], it is interesting to describe how the surface structure may be modified because of the presence of the FeTPP molecule. The investigation of the atomic displacement of the first two layers of the semi-insulating layer (Ca and F) indicates that only the atoms of the stripe on which the FeTPP molecule is straddled are faintly impacted by the adsorbate, whereas the surrounding of the molecule is not. For all the considered atoms, the displacements are less than ± 0.15 Å in both the [110] and [1-10] directions. For the vertical direction, i.e. perpendicular to the surface, it mainly occurs at the two substitutional Ca atoms having vertical displacements of ~ 0.25 Å. Additional information concerning these displacement is provided in Figs. S10.

It is now crucial to determine if the considered semi-insulating ribbon of stripes are able to correctly electronically decouple the FeTPP adsorbate with a good efficiency. The dI/dV curves acquired at three different parts of the molecule shown in Fig. 6a exhibit flat signals in the range -1.2 V to 1.5 V. Small peaks of density of states measured at -1.6 V and +1.6 V, at the pyrroles locations (red dot in the inset of Fig. 6a) indicate a probable HOMO-LUMO gap of 3.2 eV. The two dI/dV curves acquired at the phenyls and central part of the molecule are rather similar and show an occupied DOS peak at -2.4 V below the surface Fermi level whereas the unoccupied states can progressively span up to 2.5 V. These experimental data can be compared to the calculated PDOS at specific areas of the molecule as illustrated in Figs. 6b and 6c. When the PDOS includes only the phenyl groups of the FeTPP molecule (Fig. 6b), one can see two mains PDOS peaks at -2.4 eV and 2.8 eV with two weaker satellites peaks at -1.1 eV and 0.9 eV. Surprisingly, the PDOS curves calculated at the four pyrroles of the FeTPP molecule (Fig. 6c) exhibit DOS peaks at -1.1 eV and 0.9 eV, i.e. at the same energy of the satellite PDOS peaks observed in Fig. 6b. The dI/dV peaks observed at the pyrroles locations on the FeTPP at ± 1.6 V can be related with the calculated PDOS satellite peaks in Figs. 6b and 6c. Their relative weak dI/dV intensities also reveal that the insulating layer may hinder the accurate measurement of the frontier orbitals HOMO-LUMO gap of the molecule.

Generally, when adsorbates are interacting with surfaces the HOMO-LUMO gap energy of the molecules known in gas phase can be significantly reduced[44], especially if chemisorption occurs. On insulating layers, another effect can take place, arising from the potential loss at each interfaces of the double STM junction[45], involving higher applied voltages to reach frontier orbitals. Looking at the symmetrical position of the dI/dV peaks in Fig. 6a and PDOS in Fig. 6b compared to the Fermi level energy, our results indicate that the electrochemical potential of the adsorbed FeTPP molecule is weakly perturbed by the insulating layer. It also shows that the total voltage drops at each STM junction (i.e. silicon-molecule and molecule-tip) in the range of 0.4 eV. The FeTPP molecule orbitals energies calculated in gas phase are reported in Fig. 6d and clearly indicate that a group of orbitals (N° 169 to 176) having almost degenerated energies levels are the HOMO - LUMO frontiers orbitals as measured in the dI/dV and PDOS curves in Figs. 6a and 6b. These orbitals mainly involve the central macrocycle and especially the pyrroles groups of the porphyrin. Coherently, the phenyl rings of the FeTPP molecule are involved at orbitals of higher or lower energies.

## 4. CONCLUSION

In summary, we have performed the first complete experimental and theoretical investigation of the structural and electronic description of $CaF_2$ ribbons of stripes at the nanoscale. Our work shows that the initial formation of a wetting layer on the Si(100) surface is the first stage of the growth of ribbons of $CaF_2$ stripes. A detailed analysis of the STM images acquired at low temperature (9K) provides a clear description of the relative positioning of the additional $CaF_2$ molecules within the existing wetting layer structure. The characterization of the observed epitaxial structure is supported by precise DFT calculations of the ensuing surface geometry and compared with the acquired STM images. By using the DFT+U method, the electronic structure of the surface gap energy of a ribbon of stripe reproduces the one measured experimentally. In particular, we have shown that the effective charges on the Ca and F atoms of the insulating layer are similar to those determined on bulk $CaF_2$ surface reconstruction. Thus, our findings imply that the electronic insulating properties are efficient from one single monolayer. To further test the electronic properties of the $CaF_2$ insulating layer, we have studied the interactions of a single FeTPP molecule with the surface structure as it appears on the STM images. Weak interactions between the surface and the adsorbate are observed via the study of the LDOS difference between the molecule and the surface. A detailed analysis of the charge state of the adsorbed FeTPP molecule indicates typical van der Waals interactions. The electronic properties of the FeTPP molecule is described via dI/dV tunnel spectroscopy measurements acquired at several strategic locations. The comparison of these data with ad-hoc calculations of the FeTPP molecule on the surface and in gas phase further confirms that the electronic structure of the adsorbate is almost not affected by the $CaF_2$ insulating layer. The ensemble of these findings endorses that the electronic charge transfer between the adsorbate and the semi-insulating layer is weak[46]. It also demonstrates the clear insulating properties of the described $CaF_2$ ribbon of stripes. Such insulating layer on the Si(100) surface will definitively open new perspectives in the control of electronic, optic and magnetic devices at the nanoscale.

## ASSOCIATED CONTENT

**Supporting information**

The supporting information is available and attached to the submission of this manuscript as the file CaF2_analysis_SUPPL_material.pdf

## AUTHOR INFORMATION

**Corresponding author:**
Email: damien.riedel@u-psud.fr,  ORCID: 0000-0002-7656-5409

Eric Duverger: ORCID N° 0000-0002-7777-8561

Anne-Gaëlle Boyer, No known ORCID number (student)

Hélène Sauriat-Dorizon : ORCID N° Not known

Philippe Sonnet, ORCID N° 0000-0001-5891-5500

Régis Stephan, ORCID N° 0000-0003-0345-1387

Marie-Christine Hanf, ORCID N° 0000-0003-1820-504X

**Author contributions:**

All the authors participate to the writing of the manuscript. DR performed the experimental measurements and analysis as well as the DFT simulations in gas phase of the FeTPP molecule. AGB and HD participate to the molecular selection and the general discussions. ED, RS, MCH and PS performed the DFT simulations on the large atomic slab with or without molecules.

**ACNOWLEDGEMENTS**

All the authors wish to thank the French National Agency for Research (ANR) for their financial support of the CHACRA project under the contract number ANR-18-CE30-0004-01. DR would like to thank Dr. Henry Pinto from Yachay Tech. University, Ecuador, for valuable discussions. This work was granted for access to the HPC resources of the Mesocentre of Strasbourg and by the Mesocentre of Université de Franche Comté.

**FIGURES CAPTIONS:**

**Figure 1:** (a) Recall of the three dimensional balls and sticks structure of the wetting layer (WL) unit-cell following the epitaxy of Si(100) with $CaF_2$ molecules at low coverage. (b) 21.4 x 22.7 nm² STM topography (I = 32 pA, V = -2.5 V) after the epitaxy of the bare Si(100) surface with a coverage of 1.3 monolayer of $CaF_2$. The positions of the WL unit-cells (blue rectangles) are indicated in (b) as compared to the ribbon of stripes and the residual Si(100) surface areas.

**Figure 2:** (a) Three dimensional representation (perspective) of the stripes structure of a small section (19 x 41 Å²) of the topography in Fig. 1b. The vertical axis relative scale (i.e. corrugation) is increased

by software to improve the observation of the stripes structure. (b) Details of a three dimensional representation of a single stripe (top view) showing the chevrons structure. (c) 8.2 x 4.3 nm² STM topography (I = 17 pA, V = -2.5 V) showing the top view of a ribbon of stripes located nearby a bare silicon surface area. The lateral blue and red curves represent the relative height profile acquired either along a stripe or across the silicon dimer rows, respectively. The silicon dimer rows directions are symbolized by red dotted lines. Four consecutive WL unit-cells are superimposed along these red dotted lines to identify the location of the substitutional Ca atoms compared to the stripes locations.

**Figure 3:** (a) Top view of the considered silicon slab of the insulating layer electronic properties comprising two portions of $CaF_2$ stripes (four unit-cells). (b) and (c) Side A and side B of the slab, respectively, as indicated in (a). (d) Comparison of the STM topography (left panel, 35.1 x 30.8 Å², $V_s$ = -2.3 V, I = 12 pA) of two insulating stripes with the same calculated LDOS structure (right) at -1.5 eV below the Fermi energy. The positions of the Ca (green) and F (cyan) atoms from the calculated structure are reported in the left panel for comparison.

**Figure 4:** (a) Semi-logarithmic representation of dI/dV spectroscopic curve resulting of the average of ten measurements acquired at twenty different points along various stripes of the same ribbon. (b) Evolution of the calculated surface gap energy as a function of the Hubbard parameter $U_{eff.}$. (c) Calculated PDOS curves for the F, Si and Ca atoms and for the total system ($U_{eff}$ = 26.45 eV). (d) Example of a semi-logarithmic representation of the total PDOS from which a calculated surface gap energy is extracted ($U_{eff}$ = 26.45 eV).

**Figure 5:** (a) 7.7 x 3.9 nm² STM topography (I = 12 pA, V = -2.7 V) of a FeTTP molecule adsorbed on a ribbon of stripes. (b) and (c) top and side views of the FeTPP molecular structure on the insulating layer after relaxation. (d) Calculated STM image of an adsorbed FeTPP molecule (LDOS) following the conformation shown in (b) and (d). (e) and (f) Calculated spatial LDOS difference before and after adsorption of the FeTPP molecule onto a CaF$_2$ stripe. The red and blue isodensities stand for positive or negative LDOS differences, respectively.

**Figure 6:** (a) Averaged measured dI/dV curves acquired at 3 different locations on top of the FeTPP molecule. (b) and (c) Calculated PDOS curves over the four phenyl rings and pyrroles of the adsorbed FeTPP molecule, respectively. (d) Calculated molecular orbitals energies of the FeTPP in the gas phase and estimation of the HOMO-LUMO gap energy.